\documentclass[preprint,showpacs,aps]{revtex4}
\usepackage{epsf}
\usepackage{dcolumn}
\usepackage{bm}
\everymath{\displaystyle}
\begin{document}

\title{CLASSICAL AND QUANTUM SPINS IN CURVED SPACETIMES}
\author{Alexander J. Silenko}\affiliation{Institute of Nuclear Problems, Belarusian State
University, Minsk 220030, Belarus}

\date{\today}

\begin{abstract}
A comparative analysis of the Mathisson-Papapetrou and
Pomeransky-Khriplovich equations is presented. Motion of spinning
particles and their spins in gravitational fields and noninertial
frames is considered. The angular velocity of spin precession
defined by the Pomeransky-Khriplovich equations depends on the
choice of the tetrad. The connection of such a dependence with the
Thomas precession is established. General properties of spin
interactions with gravitational fields are discussed. It is shown
that dynamics of classical and quantum spins in curved spacetimes
is identical. A manifestation of the equivalence principle in an
evolution of the helicity is analyzed.
\end{abstract}
\pacs {04.20.Cv, 04.60.-m, 95.30.Sf}
\maketitle

\section{Introduction}

Spin dynamics in curved spacetimes is an important part of spin
physics. Spin effects in gravitational fields and noninertial
frames are important not only for particles but also for
gyroscopes and even celestial bodies. Many such effects can be
discovered and investigated in cosmic experiments. Therefore, a
necessary theoretical description of the spin dynamics in curved
spacetimes should be carries out.

Pioneering calculations of the spin effects in gravitational
fields were made soon after the creation of the general relativity
\cite{deS,Fokker,LT}. However, an investigation of mutual
influence of particle and spin motion in curved spacetimes was
started from the excellent work by M. Mathisson \cite{Mathisson}.
Another investigation of this problem was performed by Pomeransky
and Khriplovich \cite{PK}.

We present a comparative analysis of different equations of motion
of spinning particles and their spins, discuss their connection
with the equivalence principle, and investigate specific effects.
In the next section, we introduce the Mathisson-Papapetrou
equations (MPE). In Sec. 3, we briefly discuss general properties
of spin interactions with gravitational fields. Next section is
devoted to the comparison between the MPE and
Pomeransky-Khriplovich equations (PKE). Equations of spin motion
in stationary spacetimes are discussed in Sec. 5. In Sec. 6, the
spin effects in classical and quantum gravity are compared. A
manifestation of the equivalence principle in an evolution of the
helicity in gravitational fields and noninertial frames is
analyzed in Sec. 7. Finally, in Sec. 8 we discuss previously
obtained results and summarize the main results of the work.

Throughout the work tetrad indices are designated by first Latin
letters. Greek indices and other Latin indices run 0,1,2,3 and
1,2,3, respectively. The metric signature $(+,-,-,-)$ is chosen.
We use the designations $[\dots,\dots]$ and $\{\dots,\dots\}$ for
commutators and anticommutators, respectively. We use the term ``tetrad
vector'' for 
vectors formed from tetrad components.

\section{Mathisson-Papapetrou equations}

The famous MPE first found by Mat\-his\-son \cite{Mathisson} and
then rediscovered by Papapetrou \cite{Papapetrou} describe
dynamics of a classical spinning particle and their spin in curved
spacetimes. All multipole moments of an extended body in a
gravitational field was taken into account by Dixon \cite{Dixon}.
The explicit form of the MPE is
\begin{equation}
\begin{array}{c} \frac{Dp^\mu}{d s}=-\frac12
R_{\nu\alpha\beta}^\mu u^\nu S^{\alpha\beta}, \end{array}
\label{MPE1}
\end{equation}
\begin{equation}
\frac{DS^{\mu\nu}}{d s}=p^\mu u^{\nu}-p^\nu u^{\mu},
\label{MPE2}
\end{equation}
where $u^\nu$ and $p^\nu$ are the four-velocity and four-momentum
of the spinning particle, respectively,
$R_{\nu\alpha\beta}^\mu$ is the Riemann curvature tensor of the
spacetime, and $D/(ds)$ means the covariant derivative with
respect to the interval $ds$.

These equations should be supplemented with the condition
\cite{Mathisson,Pirani} \begin{equation} S^{\mu\nu}u_\nu=0
\label{MP}
\end{equation}
or \cite{Dixon,T,Dixon2}
\begin{equation} S^{\mu\nu}p_\nu=0.
\label{TD}
\end{equation}

The MPE characterize the pole-dipole approximation, when multipole
moments of higher orders are neglected. These equations predict
the mutual influence of particle and spin motion. In particular,
spinning particles undergo an additional force which is similar to
the Stern-Gerlach force in electrodynamics. As a result, spinning
particles do not move on geodesics in curved spacetimes.

In a zero approximation, one can neglect the mutual influence of
particle and spin motion. In this approximation, the spin tensor
is parallel transported in the spacetime and the MPE take the form
\begin{equation}
\begin{array}{c} \frac{Dp^\mu}{d s}=0,
\end{array} \label{MPA1}
\end{equation}
\begin{equation}
\begin{array}{c}
\frac{DS^{\mu\nu}}{d s}=0,
\end{array} \label{MPA2}
\end{equation}\begin{equation}
\begin{array}{c} p^\mu=mcu^\mu,
\end{array} \label{MPA3}
\end{equation}
where $m$ is the mass of the particle.

\section{General properties of spin interactions with gravitational fields}

General properties of interactions of classical spin with
gravitational fields can be obtained, when the mutual influence of
particle and spin motion is neglected. 

The curvature of spacetime conditions a precession of moving
spinning particles and gyroscopes (geodetic effect
\cite{deS,Fokker}). Additional rotation of the spin in a
gravitational field of a rotating body is caused by the frame
dragging (Lense-Thirring effect \cite{LT}). This effect results in
appearing an additional acceleration similar to the Coriolis one
and an additional precession of satellite orbits and the spin. Similar
effects take place in a
rotating frame. In the nonrelativistic approximation, resulting
motion of the spin is given by \cite{Schiff}
\begin{equation} \frac{d\bm S}{dt}=\bm\Omega\times\bm S, ~~~
\bm\Omega=\frac{3GM}{2c^2r^3}(\bm r\times\bm v)+
\frac{G}{c^2r^3}\left[\frac{3(\bm J\cdot\bm r)\bm r}{r^2}-\bm
J\right], \label{Schiff} \end{equation} where $M$ and $\bm J$ are
the mass and angular moment of the central body and $\bm v$ is the
velocity of spinning test particle. As was mentioned in Ref.
\cite{Schiff}, Eq. (\ref{Schiff}) is consistent with
approximate Mathisson-Papapetrou equation (\ref{MPA2}).

Eq. (\ref{Schiff}) results in the conclusion that an anomalous
gravitomagnetic moment (AGM) and a gravitoelectric dipole moment
(GDM) are equal to zero. Indeed, the angular velocity of spin
rotation depends on neither the mass nor the spin.
Therefore, the relation between the torque $d\bm S/dt$ and the
spin $\bm S$ is the same for all particles/gyroscopes. This is an
explicit manifestation of the absence of the AGM. The equality of angular
velocities of spin rotation of all particles in the stationary
spacetime created by the rotating body is another
manifestation of the absence of the AGM. Evidently, the latter
property is also valid for spinning particles (gyroscopes) in the
rotating frame. The absence of the GDM results from the fact that
the spin of particle at rest does not interact with a static
gravitational field if tidal forces can be neglected.

It is not evident whether the above mentioned conclusion remains
valid for quantum particles. This problem was much-discussed (see
Ref. \cite{Mashhoon2} and references therein). Nevertheless, an
explicit solution of the problem was obtained many years ago by
Kobzarev and Okun \cite{KO}. In this work, gravitational
interactions of spin-1/2 particles have been considered and
important relations for gravitational form factors at zero
momentum transfer have been derived. It was proved that
gravitational and inertial masses are equal ($f_1=g_1=1$), any
gravitomagnetic moment is ``normal" [i.e. the AGM is equal to
zero ($f_2=1$)], and the GDM is equal to zero ($f_3=0$).
The generalization of these properties to arbitrary-spin particles
was made by Teryaev \cite{T1}.

The absence of the GDM results in the absence of spin-gravity
coupling
\begin{equation} W\sim\bm g\cdot\bm S,
\label{coupling} \end{equation} where $\bm g$ is the
gravitational acceleration.

The relations obtained by Kobzarev and Okun lead to equal
frequencies of precession of quantum (spin) and classical
(orbital) angular momenta and the preservation of helicity of
Dirac particles in gravitomagnetic fields (i.e. the fields defined
by the components $g_{i0}$ of the metric tensor, see Ref.
\cite{T1} and references therein). Any reference frame
characterized by the nonzero $g_{i0}$ possesses these properties.
In particular, one can mention the gravitational field of massive
rotating body and the noninertial rotating frame.

Thus, the general properties of spin interactions with
gravitational fields are the same for classical and quantum
particles.

Similarity of Eqs. (\ref{MPA1}) and (\ref{MPA2}) conditions
conformity of particle and spin dynamics in the general
relativity. The equality of angular velocities of spin rotation of
all particles is similar to the independence of particle
accelerations in curved spacetimes on their masses. Therefore, the
above discussed general properties of spin interactions with
gravitational fields can be considered as the manifestations of
the equivalence principle in spin-gravity interactions (see Ref.
\cite{T1}).

\section{Comparison between Mathisson-Papapetrou and Pomeransky-Khriplovich equations}

There are two possible methods used for the derivation of the MPE
and PKE \cite{PK}. First method consists in a search for
appropriate covariant equations. This method was utilized for the
derivation of classical equations of spin motion in
electromagnetic fields \cite{Thoms,BMT,GN}. The
Thomas-Bargmann-Michel-Telegdi (T-BMT) equation \cite{Thoms,BMT}
and the Good-Nyborg equation (GNE) \cite{GN} describe spin
dynamics in uniform and nonuniform electromagnetic fields,
respectively. The same method was applied by Mat\-his\-son
\cite{Mathisson} and Papapetrou \cite{Papapetrou} for obtaining
equations of spin motion in curved spacetimes.

Second method consists in a deduction of equations on the basis of
some physical principles without an attempt to obtain covariant
final equations. Pomeransky and Khriplovich \cite{PK} used this
method for the derivation of equations of spin motion in
electromagnetic fields with allowance for terms of the first and
second orders in spin. This method is based on the fact that the
three-component spin defined in a particle rest frame is a
noncovariant quantity \cite{PK}. The use of covariant equations
may be possible if coordinates are redefined \cite{PK}. The
validity of the ``noncovariant'' \cite{PK} approach was confirmed
with a comparison between the GNE, PKE for the electromagnetic
field \cite{PK}, and the equation deduced in Refs.
\cite{JETP,UNTSMNE} from the Hamiltonian for
spin-1 particles in the electromagnetic
field \cite{YB}. It was shown \cite{JETP,UNTSMNE} that the
Foldy-Wouthyusen (FW) transformation followed by the
semiclassical transition results in the equation of spin motion
which agrees with the PKE but contradicts to the GNE. This is an
indirect confirmation of the noncovariant approach which was also
used \cite{PK} for description of gravitational interactions.

To find a connection between MPE and PKE, we use the results
obtained by Chicone, Mashhoon, and Punsly \cite{CMP}. The
relation between the four-momentum and four-velocity has the form
\begin{equation} p^\mu=mcu^\mu+E^\mu,
\label{eqCMP} \end{equation} where the order of magnitude of
$E^\mu$ is given by
\begin{equation} E^\mu\sim \frac{1}{mc}S^{\mu\nu}\frac{Dp_\nu}{ds}.
\label{ECMP} \end{equation} The additional four-force is $DE^\mu/d\tau$,
where $\tau$ is the proper time. This four-force is of the second order
in the spin \cite{CMP}. The
approximate equation of the first order in the spin resulting from
Eq. (\ref{MPE1}) is \cite{CMP}
\begin{equation}
\begin{array}{c} mc\frac{Du^\mu}{d s}=-\frac12
R_{\nu\alpha\beta}^\mu u^\nu S^{\alpha\beta}.
\end{array} \label{MPCMP}
\end{equation}

Eq. (\ref{Schiff}) and the PKE unambiguously show that the spin
dynamics depends on derivatives of the metric tensor. The right
hand side of Eq. (\ref{MPE2}) defined by 
Eqs. (\ref{MPE1}),(\ref{eqCMP}),(\ref{ECMP}) is of order of
\begin{equation}
p^\mu u^{\nu}-p^\nu u^{\mu}\sim
\frac{1}{mc}R_{\lambda\alpha\beta\gamma} u^\alpha u^{\nu}
S^{\mu\lambda}S^{\beta\gamma}.\label{Delta}
\end{equation} This quantity is much less than
terms defining the spin motion in the PKE \cite{PK}, when the
weak-field approximation is used. In this approximation
\begin{equation}|g_{\mu\nu}-\eta_{\mu\nu}|=|h_{\mu\nu}|\ll1,
\label{eqwfa}\end{equation} where the tensor
$\eta_{\mu\nu}$ characterizes the Minkowski spacetime. In
addition, the right hand side in Eq. (\ref{Delta}) is of the
second order in the spin tensor. Therefore, the correction to Eq.
(\ref{MPA2}) is rather small.
When terms of the first order in the spin are retained, the MPE
reduce to Eqs. (\ref{MPA2}),(\ref{MPCMP}).

To derive the corresponding equation for the spin (pseudo)vector $S^{\mu}$
in the same approximation, we can use the known connection between
the spin vector and the spin tensor \cite{MashhoonJMP} and Eq.
(\ref{MPCMP}). When only terms of the first order in the spin are taken
into account, the equation for the spin vector is given by
\begin{equation}
\begin{array}{c}
\frac{DS^{\mu}}{ds}=0.
\end{array} \label{eqPK} \end{equation}

This is the initial equation used by Pomeransky and Khriplovich
\cite{PK}. Therefore, we can conclude that the spin dynamics
predicted by the MPE and PKE is the same in the first-order
approximation in the spin. A possible difference between two
approaches can be caused by second-order terms in the spin
(including quadrupole interactions). Such terms was calculated in
Ref. \cite{PK} in the framework of quantum theory. In the present
work, we confine ourselves to the discussion of first-order spin
effects.

Eq. (\ref{eqPK}) should be supplemented with the orthogonality
condition
\begin{equation} S^{\mu}u_\mu=0.
\label{MPPK}
\end{equation}

The method developed by Pomeransky and Khriplovich \cite{PK} is
based on the equations of motion of particles and their spins in
the zero approximation [Eqs. (\ref{MPA1}) and (\ref{eqPK}),
respectively]. In Ref. \cite{PK}, the former equation was written
for the four-velocity and the tetrad formalism was used. The
equations for the tetrad components of the four-velocity
$u^{a}=e^a_{\mu}u^{\mu}$ and the four-component spin
$S^{a}=e^a_{\mu}S^{\mu}$ are \cite{PK}
\begin{equation}
\begin{array}{c} \frac{du_a}{d s}=\gamma_{abc}u^bu^c, \end{array}
\label{PKtv}
\end{equation}
\begin{equation}
\frac{dS_a}{d s}=\gamma_{abc}S^bu^c. \label{PKts}
\end{equation}
Here $e^a_{\mu}$ is the vierbein and
$$\gamma_{abc}=e_{a\mu;\nu}e_b^{\mu}e_c^{\nu}=-\gamma_{bac}$$
are the Ricci rotation coefficients \cite{LL}.

Evidently, these equations are similar to the equations of motion
of particles with zero anomalous magnetic moment $(g=2)$ and their
spins in an electromagnetic field:
\begin{equation}
\begin{array}{c} \frac{du_\mu}{d \tau}=\frac{e}{mc}F_{\mu\nu}u^\nu, ~~~
\frac{dS_\mu}{d \tau}=\frac{e}{mc}F_{\mu\nu}S^\nu, \end{array}
\label{ef}
\end{equation}
where $F_{\mu\nu}$ is the electromagnetic field tensor.

Therefore, the following correspondence takes place \cite{PK}:
\begin{equation}
\begin{array}{c} \frac{e}{mc^2}F_{ab}\leftrightarrow\gamma_{abc}u^c. \end{array}
\label{eflra}
\end{equation}

The antisymmetric electromagnetic field tensor has six independent
components and is defined by the electric and magnetic fields:
\begin{equation}
\begin{array}{c} F_{ab}
\leftrightarrow (\bm{E},\bm{B}).
\end{array} \label{efmi}
\end{equation}
One can similarly define the gravitoelectric and gravitomagnetic
fields:
\begin{equation}
\begin{array}{c} \frac{e}{mc}\bm{E}\leftrightarrow\bm{\mathcal{E}},
~~~\frac{e}{mc}\bm{B}\leftrightarrow\bm{\mathcal{B}},
~~~c\gamma_{abc}u^c
\leftrightarrow(\bm{\mathcal{E}},\bm{\mathcal{B}}).
\end{array} \label{gefmi}
\end{equation}
An important difference between the electromagnetic and
gravitational interactions consists in the fact that
$\gamma_{abc}u^c$ is not a tensor. Explicit expressions for the
gravitoelectric and gravitomagnetic fields are (see Ref.
\cite{PK})
\begin{equation}
\begin{array}{c} \mathcal{E}_{\hat{i}}=c\gamma_{0ic}u^{c},~~~
\mathcal{B}_{\hat{i}}=-\frac{c}{2}e_{\hat{i}\hat{k}\hat{l}}\gamma_{klc}u^{c},
\end{array} \label{expl}
\end{equation}
where $e_{\hat{i}\hat{k}\hat{l}}$ is the antisymmetric tensor with
spatial components. To avoid misleading coincidences, zero and
spatial tetrad indexes are marked with hats (except for the Ricci
rotation coefficients).

The comparison with the T-BMT equation \cite{Thoms,BMT} allows to obtain
the angular velocity of spin precession. Pomeransky and
Khriplovich introduced the three-component spin (pseudo)vector
$\bm S$ and obtained the exact equation of its precession
\cite{PK}
\begin{equation}
\begin{array}{c} \frac{d\bm S}{dt}=\bm\Omega\times\bm S,~~~
\Omega_{\hat{i}}=ce_{\hat{i}\hat{k}\hat{l}}\left(\frac12\gamma_{klc}
+\frac{u^{\hat{k}}}{u^{\hat{0}}+1}\gamma_{0lc}
\right)\frac{u^{c}}{u^0}
\end{array} \label{omega}
\end{equation}
that is equivalent to
\begin{equation}
\begin{array}{c} \bm{\Omega}=\frac{1}{u^0}\left[-\hat{\bm{\mathcal{B}}}
+\frac{\hat{{\bm u}}\times\hat{\bm{\mathcal{E}}}}
{u^{\hat{0}}+1}\right].
\end{array} \label{omgem}
\end{equation}

The factor $1/u^0$ is caused by the transition to the
differentiation with respect to the world time $t$. The
definitions of $\bm\Omega$ in Refs. \cite{PK,Obzor} and the
present work differ in sign. When the differentiation is performed
with respect to the tetrad time ($d\hat{t}=u^{\hat{0}}dt/u^{0}$),
Eq. (\ref{omgem}) coincides with the T-BMT equation for Dirac
particles ($g=2$). The gravitoelectric and gravitomagnetic fields
are defined via their tetrad components. The quantity $\bm\Omega$
characterizes the spin precession in the world frame, while the
spin $\bm S$ is defined in the particle rest frame. In this
connection, the dependence of $\bm\Omega$ on the choice of the
tetrad must result from a change of the particle rest frame.

For a Schwarzschild metric, the exact expression for $\bm{\Omega}$ was obtained in Ref.
\cite{Obzor}.

The corresponding equation of particle motion has the form
\begin{equation}
\begin{array}{c} \frac{d\hat{\bm{u}}}{dt}=\frac{u^{\hat{0}}}{u^0}\left(\hat{\bm{\mathcal{E}}}+
\frac{\hat{{\bm
u}}\times\hat{\bm{\mathcal{B}}}}{u^{\hat{0}}}\right),  ~~~ \frac{d{u}^{\hat{0}}}{dt}=
\frac{\hat{\bm{\mathcal{E}}}\cdot\hat{{\bm
u}}}{u^0}.
\end{array} \label{force}
\end{equation}
When the differentiation is performed with respect to the tetrad
time, Eq. (\ref{force}) coincides with the Lorentz equation. Eqs.
(\ref{PKtv}),(\ref{force}) describe the particle motion along
geodesic lines.

Definition (\ref{expl}) of the gravitoelectric and gravitomagnetic
fields significantly differs from the usual one. In particular,
the Pomeransky-Khriplovich gravitomagnetic field is nonzero even
for a static metric.

There is a one-to-one correspondence between the angular velocity of
precession of the three-component spin and spin-dependent terms in
classical \cite{PK} and quantum \cite{BLP} Lagrangians and Hamiltonians.
To derive spin-dependent corrections to classical Lagrangians, Poisson
brackets was used in Ref. \cite{PK}. When classical and quantum expressions
for $\bm{\Omega}$ coincide, the spin-dependent terms in classical and
quantum Lagrangians/Hamiltonians derived with the Poisson brackets and
commutators, respectively, are also the same. These terms are given by
\begin{equation}
\begin{array}{c} {\mathcal{L}}={\mathcal{L}}_0+\bm\Omega\cdot\bm
S,~~~{\cal{H}}={\cal{H}}_0-\bm\Omega\cdot\bm S,
\end{array} \label{LH}
\end{equation}
where ${\mathcal{L}}_0$ and ${\cal{H}}_0$ define sums of
spin-independent terms. It will be shown below that Eq.
(\ref{omega}) agrees with corresponding equations derived in the
framework of quantum theory. As a result, classical Lagrangians and
Hamiltonians defined by Eqs. (\ref{omega}) and (\ref{LH}) must agree with
corresponding quantum Hamiltonians. Thus, the PKE are consistent
with the quantum gravity at least in the first-order approximation
in the spin.

Influences of the spin on a particle trajectory in a gravitational
field predicted by the MPE and PKE significantly differ
\cite{PK,Obzor}. It was stated in Refs. \cite{PK,Obzor,KP} that
the MPE are not consistent with Eq. (\ref{Schiff}) describing the
geodetic effect (gravitational spin-orbit interaction). In the
Pomeransky-Khriplovich approach, the consistence of motion of
particles and their spins results from Eq. (\ref{LH}).

\section{Equations of spin motion in stationary spacetimes}

While MPE (\ref{MPA2}) is equivalent to PKE (\ref{eqPK}), general
equation of spin motion (\ref{omega}) was obtained only in the
framework of the Pomeransky-Khriplovich approach. However, the
Pomeransky-Khriplovich method needs to be gro\-un\-ded. Eqs.
(\ref{PKts}) and (\ref{ef}) describing the motion of the
four-component spin vector in gravitational and electromagnetic
fields, respectively, are very similar. However, there exists an
important difference between the corresponding equations for the
three-component spin vector. Since the latter vector is defined in
the particle rest frame, the spatial components of the
four-velocity satisfying Eq. (\ref{MPPK}) are equal to zero in
this frame. Such a definition of the velocity is ambiguous because
this quantity can be characterized by covariant, contravariant,
and tetrad vectors. A definite choice can be made due to a local
Lorentz invariance. The spacetime metric tensor locally has the
Minkowski form $\eta_{\mu\nu}$ of special relativity in any
freely-falling reference frame including the particle rest frame
(see Ref. \cite{Will}). Tetrad components of any vector are
similar to components of vectors in a flat spacetime. In
particular, covariant and contravariant tetrad components of
vectors are equal up to sign. Since the spacetime interval in
tetrad coordinates takes the Minkowski form,
tetrad reference frames are flat and correspond to local Lorentz frames.

The particle velocity is
zero and the spacetime is flat in the particle rest frame.
In this frame, just spatial tetrad components of the particle velocity
are zero ($\hat{\bm u}=0$). It is natural because any observer carries a tetrad frame
(see Ref. \cite{MTW}). Corresponding
covariant and contravariant components 
($u^i$ and $u_i$,
respectively) can be nonzero. Since the orthogonality condition
can be written in the form
\begin{equation} S^{a}u_a=0,
\label{MPPKa}
\end{equation}
the three-component spin is composed by spatial components of the
four-component tetrad spin at condition that $\hat{\bm u}=0$. Thus,
the three-component spin is a tetrad (pseudo)vector. Such a
definition of the three-component spin was 
used in Refs. \cite{PK,Obzor}.

The definition of the three-component spin as a tetrad
(pseudo)vector can be additionally justified by its consistency
with the definition of the spin operator in quantum theory. The
covariant Dirac equation for spin-1/2 particles in curved
spacetimes has the form
\begin{equation}
(i\hbar\gamma^a D_a-mc)\psi=0, ~~~~~~~ a=0,1,2,3,
\label{eqin}\end{equation} where $\gamma^a$ are the Dirac
matrices. The spinor
covariant derivatives are defined by
\begin{equation}
D_a=e_a^\mu D_\mu, ~~~ D_\mu=\partial _\mu+\frac
i4\sigma_{ab}\Gamma_\mu^{ab},
\label{eqin2}\end{equation} where
$\Gamma_\mu^{ab}=-\Gamma_\mu^{ba}$ are the Lorentz
connection coefficients,
$\sigma^{ab}=i(\gamma^a\gamma^b-\gamma^b\gamma^a)/2$
(see Refs. \cite{Ob1,Ob2} and references therein). Because the matrices
$\gamma^a$ are defined in the tetrad frame, they coincide with the Dirac matrices.

To obtain the equations of motion of particles and their spins,
one can in principle use any tetrad. However, it does not mean
that a choice of the tetrad is not important. In Eq.
(\ref{omega}), the angular velocity of spin rotation should
correspond to the quantity measured by a local observer. As a
result, parameters of the definite tetrad frame carried by the
observer should be substituted into this equation. For the
observer in a uniformly accelerated, rotating frame, the tetrad
$\bm e_a$ transports along the observer's world line according to
\cite{MTW}
\begin{equation}
\frac{d\bm e_a}{d\tau}=\mathbf{\Xi} \bm e_a,
\label{ART}\end{equation}
where $\mathbf{\Xi}$ is the antisymmetric rotation tensor. This tensor
consists of the Fermi-Walker part
$\mathbf{\Xi}_{FW}$ and the spatial rotation part $\mathbf{\Xi}_{R}$ \cite{MTW}:
\begin{equation}
\Xi^{\mu\nu}=\Xi^{\mu\nu}_{FW}+\Xi^{\mu\nu}_{R},~~~
\Xi^{\mu\nu}_{FW}=a^\mu u^\nu-a^\nu u^\mu,~~~
\Xi^{\mu\nu}_{R}=u_\alpha\omega_\beta\epsilon^{\alpha\beta\mu\nu},
\label{FWR}\end{equation}
where $a^\mu$ is the four-acceleration of the observer, $\omega^\mu$ is its
four-rotation, and $\epsilon^{\alpha\beta\mu\nu}$ is the Levi-Civita tensor.

For the uniformly accelerated, rotating frame, the exact formula for the
orthonormal tetrad satisfying Eqs.
(\ref{ART}),(\ref{FWR}) was found by Hehl and Ni \cite{HN}.
The corresponding vierbein has the form
\begin{equation}
e^{\hat{0}}_{0}=1+\frac{\bm a\cdot\bm r}{c^2}, ~~~e^{\hat{0}}_{i}=0,~~~e^{\hat{i}}_{0}=-g_{0i}=
\frac{(\bm\omega\times\bm r)^{i}}{c},~~~
e^{\hat{i}}_{j}=\delta_{ij},
\label{vierbHN}\end{equation}
where $\delta_{ij}$ is the Kronecker delta. Vierbein (\ref{vierbHN}) is attributed
to the observer being at rest in the uniformly accelerated, rotating frame
\cite{MTW,HN}.

Since the equivalence principle predicts the equivalence of gravitational fields and
noninertial frames, the result obtained in Ref. \cite{HN} can be used for any
spacetime defined by a 
metric tensor with $g_{0i}\neq 0$. Nonzero $g_{0i}$ components are connected
with the proper local three-rotation $\bm\omega$.
In the weak-field approximation, the generalization of the tetrad
found in Ref. \cite{HN} is given by
\begin{equation}
e^{\hat{0}}_{0}=1+\frac{1}{2}h_{00}, ~~~e^{\hat{0}}_{i}=0,~~~e^{\hat{i}}_{0}=-g_{0i},~~~
e^{\hat{i}}_{j}=\delta_{ij}-\frac{1}{2}h_{ij}.
\label{vier1}\end{equation}
This vierbein can also be presented in the
equivalent forms
\begin{equation}
e_{\hat{0}0}=1+\frac{1}{2}h_{00}, ~~~
e_{\hat{0}i}=0,~~~e_{\hat{i}0}=g_{0i},~~~
e_{\hat{i}j}=-\delta_{ij}+\frac{1}{2}h_{ij}
\label{vier2}\end{equation}
and
\begin{equation}
e_{\hat{0}}^{0}=1-\frac{1}{2}h_{00}, ~~~
e_{\hat{0}}^{i}=g_{0i},~~~e_{\hat{i}}^{0}=0,~~~
e_{\hat{i}}^{j}=\delta_{ij}+\frac{1}{2}h_{ij}.
\label{vier3}\end{equation}

Vierbeins (\ref{vier1})--(\ref{vier3}) are connected with the observer
at rest.

The nonzero Ricci rotation coefficients are equal to
\begin{equation}\begin{array}{c}
\gamma_{i00}=\frac{1}{2}g_{00,i}=-\gamma_{0i0},~~~
\gamma_{i0j}=\frac{1}{2}(g_{0i,j}+g_{0j,i})=-\gamma_{0ij},\\
\gamma_{ij0}=\frac{1}{2}(g_{0j,i}-g_{0i,j}),~~~
\gamma_{ijk}=\frac{1}{2}(g_{jk,i}-g_{ik,j}).
\end{array}
\label{Ricci}\end{equation}

Alternatively, one can use the vierbeins proposed by Pomeransky and
Khriplovich \cite{PK}
\begin{equation}
e^{\hat{0}}_{0}=1+\frac{1}{2}h_{00}, ~~~e^{\hat{0}}_{i}=\frac{1}{2}g_{0i},
~~~e^{\hat{i}}_{0}=-\frac{1}{2}g_{0i},~~~
e^{\hat{i}}_{j}=\delta_{ij}-\frac{1}{2}h_{ij},
\label{vPK}\end{equation}
and Landau and Lifshitz \cite{LL}
\begin{equation}
e^{\hat{0}}_{0}=1+\frac{1}{2}h_{00}, ~~~e^{\hat{0}}_{i}=g_{0i},~~~
e^{\hat{i}}_{0}=0,~~~
e^{\hat{i}}_{j}=\delta_{ij}-\frac{1}{2}h_{ij}.
\label{vLL}\end{equation}
Formulae (\ref{vier1})--(\ref{vLL}) are given in the weak-field approximation.

The expression for the
Ricci rotation coefficients obtained in Ref. \cite{PK} with the
Pomeransky-Khriplovich tetrad differs from Eq. (\ref{Ricci}):
\begin{equation}\begin{array}{c}
\gamma_{abc}=\frac{1}{2}(h_{bc,a}-h_{ac,b})=\frac{1}{2}(g_{bc,a}-g_{ac,b}).
\end{array}
\label{RPK}\end{equation}

If a tetrad does not satisfy Eqs. (\ref{ART}),(\ref{FWR}), it is
not attributed to the observer's frame. We can consider an
influence of the choice of the tetrad on equation of spin motion
(\ref{omega}).
The connection between different tetrad frames can be
expressed by appropriate Lorentz transformations. Let the vierbeins $e_a^\mu$ and
${e'}_a^\mu$ define two tetrad frames and the unprimed vierbein is
attributed to the observer's rest frame.
The connection between tetrad and world coordinates is given by
\begin{equation}
dx^a=e^a_\mu dx^\mu, ~~~ d{x'}^a={e'}^a_\mu dx^\mu.
\label{eqin3}\end{equation}
Eq. (\ref{eqin3}) leads to the relationship between tetrad coordinates in
two frames:
\begin{equation}
d{x}^a=T^a_b d{x'}^b, ~~~ T^a_b={e}^a_\mu {e'}^\mu_b.
\label{relationsh}\end{equation}

Since this relationship defines a Lorentz
transformation, the primed frame moves with the relative velocity
$\bm V$. This velocity is equal to zero only when
$T^{\hat{0}}_{\hat{i}}=0,~T_{\hat{0}}^{\hat{i}}=0,~i=1,2,3$.
In this case, the primed frame can be obtained from the unprimed one with a turn of the
triad $\bm e_{\hat{i}}$ in the three-dimensional space.
Evidently, this turn does not change the observer's rest frame and the
condition $\bm V=0$ remains valid. Such a turn does not distort the dynamics
of particles and their spins, while it changes the connection between world and
tetrad velocities.

In the general case, the primed tetrad is attributed to the reference frame
moving with the velocity $\bm V$ relative to the observer.
As a rule, this velocity is time-dependent.
Eq. (\ref{MPPKa}) shows that the three-component spins $\bm S$ and $\bm S'$
are defined in different reference frames. Since accelerations of these
frames do not equal, the Thomas precession causes a difference between
angular velocities of rotation of the (pseudo)vectors $\bm S$ and $\bm S'$.

As a result, the 
reason of the change in the spin motion equation is the Thomas
precession. For relativistic particles, the angular velocity of
the Thomas precession is given by \cite{Thoms,Malykin}
\begin{equation}
\bm\Omega_T=-\frac{\gamma}{\gamma+1}(\bm\beta\times\dot{\bm\beta}),
\label{Thoms}\end{equation}
where $\gamma=1/\sqrt{1-\beta^2}$ is the Lorentz factor.

The dependence of spin motion equation (\ref{omega}) from the
choice of the tetrad was not taken into account in Refs.
\cite{PK,Obzor}. The tetrad used in Refs. \cite{PK,Obzor} for a
derivation of equations of spin motion in the world frame
satisfies Eqs. (\ref{ART}),(\ref{FWR}) only for static spacetimes.
To determine the observable angular velocity for nonstatic
spacetimes, one needs to supplement the PKE with the correction
for the Thomas precession.

To illustrate a dependence of Eq. (\ref{omega}) from the choice of
the tetrad, we consider the spin motion in the rotating frame.
This problem can be solved exactly. The metric tensor is given by
\cite{HN}
\begin{equation}
\begin{array}{c}
g_{\mu\nu}=\left(\begin{array}{cccc} 1-\frac{(\bm\omega\times\bm r)^2}{c^2} &
-\frac{(\bm\omega\times\bm r)^{(1)}}{c}&
 -\frac{(\bm\omega\times\bm r)^{(2)}}{c} &
 -\frac{(\bm\omega\times\bm r)^{(3)}}{c}
 \\ -\frac{(\bm\omega\times\bm r)^{(1)}}{c} & -1 & 0& 0
 \\ -\frac{(\bm\omega\times\bm r)^{(2)}}{c}  & 0 & -1& 0 \\
-\frac{(\bm\omega\times\bm r)^{(3)}}{c}& 0 & 0& -1
 \end{array}\right),
 \end{array}\label{gmn}\end{equation}
where $\bm\omega$ is the
angular frequency of the frame rotation. The use
of Eqs. (\ref{force}),(\ref{vier1})--(\ref{vier3})
results in the following equation of particle motion:
\begin{equation}
\begin{array}{c} \frac{d\hat{\bm{u}}}{dt}=-\bm\omega\times\hat{\bm{u}}, ~~~ \frac{d{u}^{\hat{0}}}{dt}=0.
\end{array} \label{fHN}
\end{equation}
Eq. (\ref{fHN}) leads to the right equation for the contravariant acceleration
$du^\mu/(dt)$ coinciding with the well-known formula \cite{Gol} for
the sum of the Coriolis and centrifugal accelerations.

The corresponding angular velocity of spin motion obtained from Eq.
(\ref{omega}) is given by
\begin{equation}
\begin{array}{c} \bm\Omega=-\bm\omega.
\end{array} \label{OHN}
\end{equation}
This formula is also exact and coincides with previous results
\cite{HN,Gor,Mashhoon}.

Pomeransky-Khriplovich vierbein (\ref{vPK}) leads to the formula
\begin{equation}
\begin{array}{c} \bm\Omega=-\bm\omega+\frac{\bm{u}\times(\bm{u}\times\bm\omega)}
{2u^0(u^0+1)}.
\end{array} \label{OPK}
\end{equation}
The use of Landau-Lifshitz vierbein (\ref{vLL}) results in
\begin{equation}
\begin{array}{c} \bm\Omega=-\bm\omega+\frac{\bm{u}\times(\bm{u}\times\bm\omega)}
{u^0(u^0+1)}.
\end{array} \label{OLL}
\end{equation}
Eqs. (\ref{OPK}),(\ref{OLL}) are obtained in the weak-field approximation.
Evidently, these equations do not give the observable angular velocity defined by
Eqs. (\ref{OHN}).

\section{Comparison of spin effects in classical and quantum gravity}

The correspondence principle formulated by Niels Bohr predicts a similarity
of classical and quantum effects.

The best compliance between the description of spin effects in
classical and quantum gravity was proved in Refs. \cite{PRD,PRD2}.
In these works, some Hamiltonians in the Dirac representation
derived in Refs. \cite{Ob1,Ob2,HN} from initial Dirac equation
(\ref{eqin}) were used. The initial Dirac Hamiltonians were
transformed to the Foldy-Wouthyusen (FW) representation by the
method elaborated in Ref. \cite{JMP}. The FW representation
\cite{FW} occupies a special place in the quantum theory.
Properties of this representation are unique. The Hamiltonian and
all operators are block-diagonal (diagonal in two spinors).
Relations between the operators in the FW representation are
similar to those between the respective classical quantities. For
relativistic particles in external fields, operators have the same
form as in the nonrelativistic quantum theory. For example, the
position operator is $\bm r$ and the momentum one is $\bm
p=-i\hbar\nabla$. These properties considerably simplify the
transition to the semiclassical description. As a result, the FW
representation provides the best possibility of obtaining a
meaningful classical limit of the relativistic quantum mechanics.
The basic advantages of the FW representation are described in
Refs. \cite{FW,CMcK,JMP}. The method of the FW transformation for
relativistic particles in external fields was proposed in Ref.
\cite{JMP}.

The exact transformation of the Dirac equation for the metric
\begin{equation}
ds^2=V^2(\bm r)(dx^0)^2-W^2(\bm r)(d\bm r\cdot\bm r)
\label{metric}\end{equation} to the Hamiltonian form was carried
out by Obukhov \cite{Ob1,Ob2} ($\hbar=c=1$):
\begin{equation}
i\frac {\partial \psi}{\partial t}={\cal H}\psi, ~~~
{\cal H} = \beta m V+\frac12\{{\cal F},\bm\alpha\cdot\bm p\},
\label{eqO}\end{equation}
where ${\cal F} = V/W$.
Hamiltonian (\ref{eqO}) covers the cases of a weak Schwarzschild field
in the isotropic coordinates
\begin{equation}
V =\! \left(1 - {\frac {GM}{2r}}\right)\left(1 + {\frac {GM} {2r}}
\right)^{-1}\!, ~~~ W =\! \left(1 + {\frac {GM}{2r}}\right)^2\!
\label{Sh}\end{equation}
and a uniformly accelerated frame
\begin{equation}
V = 1 + \bm{a}\cdot\bm{r}, ~~~ W = 1.\label{acf}\end{equation}

The relativistic FW Hamiltonian derived in Ref. \cite{PRD} has the
form
\begin{equation}
\begin{array}{c}
{\cal H}_{FW}=\beta\epsilon +\frac{\beta}{2}\left\{\frac{m^2}{\epsilon
},V-1\right\}
+\frac{\beta}{2}\left\{\frac{\bm
p^2}{\epsilon },{\cal F}-1\right\}\\
-\frac{\beta m}{4
\epsilon (\epsilon +m)}\biggl[\bm{\Sigma}\cdot(\bm\phi\times\bm p)-
\bm{\Sigma}\cdot(\bm p\times\bm\phi)+ \nabla\! \cdot\!\bm\phi\biggr]
\nonumber\\
+\frac{\beta m(2\epsilon ^3+2\epsilon ^2m+2\epsilon
m^2+m^3)}{8\epsilon ^5 (\epsilon +m)^2}(\bm p\cdot\!\nabla)(\bm
p\cdot\!\bm\phi)\\+ \frac{\beta}{4\epsilon }\left[\bm{\Sigma}\cdot(\bm
f\times\bm p)- \bm{\Sigma}\cdot(\bm p\times\bm f)+\nabla\! \cdot\!\bm
f\right]-\frac{\beta(\epsilon ^2+m^2)}{4\epsilon
^5}(\bm p\cdot\!\nabla)(\bm p\cdot\!\bm f),
\label{eq7}\end{array}\end{equation}
where $\epsilon=\sqrt{m^2+\bm p^2},~\bm\phi=\nabla V,~\bm f=\nabla{\cal F}$.

The use of the FW representation dramatically simplifies the
derivation of quantum equations. The operator equations of
momentum and spin motion obtained via commutators of the
Hamiltonian with the momentum and polarization operators take the
form \cite{PRD}
\begin{eqnarray}
\frac{d\bm p}{dt}=i[{\cal H}_{FW},\bm p]=
-\frac{\beta}{2}\left\{\frac{m^2}{\epsilon
},\bm\phi\right\}-\frac{\beta}{2}\left\{\frac{\bm p^2}{\epsilon
},\bm f\right\}
\nonumber\\
+\frac{m}{2 \epsilon (\epsilon
+m)}\nabla\bigl(\bm{\Pi}\cdot(\bm\phi\times\bm
p)\bigr)
- \frac{1}{2\epsilon } \nabla\bigl(\bm{\Pi}\cdot(\bm
f\times\bm p)\bigr)
\label{eq11}\end{eqnarray} and
\begin{eqnarray}
\frac{d\bm\Pi}{dt}=i[{\cal H}_{FW},\bm\Pi]=
\frac{m}{\epsilon (\epsilon +m)}\bm\Sigma\times\left(\bm\phi\times\bm p\right)
-\frac{1}{\epsilon }\bm\Sigma\times\left(\bm f\times\bm p\right),
\label{eq12}\end{eqnarray}
respectively.

Let us pass to the studies of semiclassical limit of these equations.
The contribution of the lower spinor is negligible and the transition to the
semiclassical description is performed
by averaging the operators in the equations for the upper spinor \cite{JMP}. It
is usually possible to neglect the
commutators between the coordinate and momentum operators. As a result, the
operators $\bm \sigma$ and $\bm p$ should be
substituted by the corresponding classical quantities: the polarization vector
(doubled average spin), $\bm \xi$, and the momentum.
For the latter quantity, we retain the notation $\bm p$. The semiclassical
equations of motion are \cite{PRD}
\begin{eqnarray}
\frac{d\bm p}{dt}= -\frac{m^2}{\epsilon }\bm\phi-\frac{\bm
p^2}{\epsilon }\bm f+\frac{m}{2 \epsilon (\epsilon
+m)}\nabla\bigl(\bm{\xi}\cdot(\bm\phi\times\bm
p)\bigr)
- \frac{1}{2\epsilon } \nabla\bigl(\bm{\xi}\cdot(\bm
f\times\bm p)\bigr) \label{eq14}\end{eqnarray} and
\begin{equation}
\frac{d\bm\xi}{dt}= \frac{m}{\epsilon (\epsilon
+m)}\bm\xi\times\left(\bm\phi\times\bm p\right)- \frac{1}{\epsilon
}\bm\xi\times\left(\bm f\times\bm p\right),
\label{eq15}\end{equation} respectively. In Eq. (\ref{eq14}), two
latter terms describe a force dependent on the spin. This force is
similar to the electromagnetic Stern-Gerlach force (see Ref.
\cite{JMP}). Because it is weak, the approximate semiclassical
equation of particle motion takes the form
\begin{equation}
\frac{d\bm p}{dt}=
-\frac{m^2}{\epsilon }\bm\phi-\frac{\bm p^2}{\epsilon }\bm f.
\label{eq16}\end{equation}
The angular velocity of spin rotation is given by
\begin{equation}
\bm\Omega=-\frac{m}{\epsilon (\epsilon +m)}\left(\bm\phi\times\bm
p\right)+\frac{1}{\epsilon}\left(\bm f\times\bm p\right).
\label{eqom}\end{equation}

We can find similar equations describing a change of the direction of particle
momentum, $\bm n=\bm p/p\;$:
\begin{equation}
\frac{d\bm n}{dt}=\bm\omega\times\bm n, ~~~ \bm\omega=
\frac{m^2}{\epsilon p}\bigl( \bm\phi\times\bm
n\bigr)+\frac{p}{\epsilon}\bigl(\bm f\times\bm n\bigr).
\label{eq18}\end{equation}

A simple calculation shows that the corresponding equations of
motion obtained from the PKE for given metric (\ref{metric})
coincide with Eqs. (\ref{eq14})--(\ref{eq18}). In particular, the
gravitational Stern-Gerlach force defined by Eq. (\ref{eq14})
coincides with that obtained from the PKE. The comparison with
previous results obtained in framework of classical gravity was
carried out in Ref. \cite{PRD}.

Although the gravitational Stern-Gerlach forces are rather weak, they are
important. These forces lead to the violation of the weak equivalence
principle due to deflections of spinning particles from geodesic lines
\cite{Plyatsko}.

Let us consider the interaction of particles
with a spherically symmetric gravitational field and compare the obtained
formulae with previous results.
This field is a weak limit of the Schwarzschild
one which yields
\begin{equation} V=1-\frac{GM}{r}, ~~~
W=1+\frac{GM}{r}.\label{eq21}\end{equation}

When we neglect the terms of order of $\frac{(\bm p\cdot\nabla)(\bm p\cdot\bm g)}{\epsilon^2}$,
Hamiltonian (\ref{eq7}) takes the form
\begin{eqnarray}
{\cal H}_{FW}=\beta\epsilon -\frac{\beta}{2}\left\{\frac{\epsilon^2+\bm p^2}{\epsilon},\frac{GM}{r}\right\}
-\frac{\beta(2\epsilon+m)}{4
\epsilon (\epsilon +m)}\biggl[2\bm{\Sigma}\cdot(\bm g\times\bm p)+ \nabla\cdot\bm g\biggr],
\label{eqd}\end{eqnarray}
where $\bm g$ is the Newtonian acceleration.

Neglecting the Stern-Gerlach force, one gets the semiclassical expressions for the
angular velocities of rotation of unit momentum vector, $\bm n=\bm p/p$,
and spin:
\begin{eqnarray}
\bm\omega=-\frac{\epsilon^2+\bm
p^2}{\epsilon\bm p^2}\bm g\times\bm p=\frac{GM}{r^3}\cdot
\frac{\epsilon^2+\bm p^2}{\epsilon\bm p^2}\bm l,
\label{eq25}\end{eqnarray}
\begin{eqnarray}
\bm\Omega=-\frac{2\epsilon
+m}{\epsilon (\epsilon +m)} \bm g\times\bm
p=\frac{GM}{r^3}\cdot\frac{2\epsilon +m}{\epsilon (\epsilon +m)}
\bm l, \label{equat}\end{eqnarray} where $\bm l=\bm r\times\bm p$
is the angular moment.

Eqs. (\ref{eq25}) and (\ref{equat}) agree with the classical
gravity. Eq. (\ref{eq25}) leads to the expression for the angle of
particle deflection by a gravitational field
\begin{equation}
\theta=\frac{2GM}{\rho}\left(2+\frac{m^2}{\bm
p^2}\right)=\frac{2GM}{\rho \bm v^2}\left(1+\bm v^2\right)
\label{eq27}\end{equation} coinciding with Eq. (13) of Problem
15.9 from Ref. \cite{AR} (see also Ref. \cite{Far}).
Eq. (\ref{equat}) coincides with the corresponding classical equation obtained in Ref. \cite{PK}. This directly
proves the full compatibility of quantum and classical
considerations.

In the nonrelativistic approximation, Eq. (\ref{equat}) coincides
with corresponding formula (\ref{Schiff}) for the classical
gyroscope. Such a similarity \cite{KO} of classical and quantum
rotators is a manifestation of the equivalence principle (see e.g.
Refs. \cite{T1,T2} and references therein). In the nonrelativistic
approximation, the last term in Hamiltonian (\ref{eqd}) describing
the spin-orbit and contact (Darwin) interactions coincides with
the corresponding term in Ref. \cite{DH}.

Performing the FW transformation for relativistic particles made
it possible to solve the problem of existence of the dipole
spin-gravity coupling in a static gravitational field \cite{PRD}.
This problem was discussed for a long time (see Refs.
\cite{Mashhoon2,Ob1,Ob2,PRD} and references therein). Evidently,
this coupling given in form (\ref{coupling}) contradicts to the
theory \cite{PK,PRD} and violates both the CP invariance and the
relation predicting the absence of the GDM \cite{KO}. The
classical and quantum approaches lead to the same conclusion.

The equation for the Hamiltonian and the equations of momentum and spin
motion derived in Ref. \cite{PRD} for a relativistic particle in a
uniformly accelerated frame agree
with the corresponding nonrelativistic expressions from \cite{HN,Huang}.
The general equations for the angular velocities of rotation of unit
momentum vector and spin are given by
\begin{equation}
\bm\omega=\frac{\epsilon}{\bm
p^2}\bigl(\bm a\times\bm p\bigr), ~~~ \bm\Omega=\frac{\bm
a\times\bm p}{\epsilon +m}. \label{eq33}\end{equation}

The FW Hamiltonian and the operators of velocity and acceleration
were also calculated for the Dirac particle in the rotating frame
\cite{PRD2}. The exact Dirac Hamiltonian derived in Ref. \cite{HN}
was used. In Ref. \cite{PRD2}, perfect agreement between classical
and quantum approaches was also established. The operators of
velocity and acceleration are equal to
\begin{eqnarray}
\bm v=\beta\frac{\bm p}{\epsilon}-\bm\omega\times\bm r, ~~~
\epsilon =\sqrt{m^2+\bm p^2},
\nonumber\\
\bm w=2\beta\frac{\bm p\times\bm\omega}{\epsilon}
+\bm\omega\times(\bm\omega\times\bm r)
=2\bm v\times\bm\omega-\bm\omega\times(\bm\omega\times\bm r).
\label{eqvn}\end{eqnarray}
Quantum formula (\ref{eqvn}) for
the acceleration of the relativistic spin-1/2 particle coincides with the
classical formula \cite{Gol} for
the sum of the Coriolis and centrifugal accelerations. Obtained results also agree
with the corresponding nonrelatiistic formulae from \cite{HN}.

Thus, the classical and quantum approaches are in full agreement.
Pu\-re\-ly quantum effects are not too important. They consist in
appearing some additional terms in the FW Hamiltonian. However,
these terms are proportional to derivatives of $\bm\phi$ and $\bm
f$ and similar to the well-known Darwin term in the
electrodynamics. As a result, their influence on the motion of
particles and their spins in gravitational fields can be
neglected. The classical and quantum equations of motion of
particles and their spins are almost identical and can differ only
in small terms.

\section{Equivalence principle and spin}

As mentioned above, the absence of the AGM and GDM is very similar to the
weak equivalence principle. All classical and quantum spins (gyroscopes)
precess with the same angular velocity, while all classical and quantum
particles move with the same acceleration. An equivalence of the inertia
and gravity manifests in the fact that all gravitational and inertial
phenomena are exhaustively defined by the metric tensor and four-velocity.

Another manifestation of the equivalence principle was found in
Ref. \cite{PRD}. It was shown that the motion of momentum and spin
differs in a static gravitational field and a uniformly
accelerated frame but the helicity evolution coincides. In Eqs.
(\ref{eqom}),(\ref{eq18}) $\bm\phi$ depends only on $g_{00}$ but
$\bm f$ is also a function of $g_{ij}$.

The spin rotates with respect to the momentum direction and the
angular velocity of this rotation is
\begin{equation}
\bm o=\bm\Omega-\bm\omega=-\frac{m}{p}\bigl(\bm\phi\times\bm
n\bigr). \label{eq19}\end{equation} The quantity $\bm o$ does not
depend on $\bm f$ and vanishes for massless particles. Therefore,
the gravitational field cannot change the helicity of massless
Dirac particles. The evolution of the helicity
$\zeta\equiv|\bm\xi_\| |=\bm\xi\cdot\bm n$ of massive particles is
defined by the formula
\begin{equation}
\frac{d\zeta}{dt}=(\bm
\Omega-\bm\omega)\cdot(\bm\xi_\bot\times\bm
n)=-\frac{m}{p}\left(\bm\xi_\bot\cdot\bm\phi\right),
\label{eq20}\end{equation}
where $\bm\xi_\bot=\bm\xi-\bm\xi_\|$.

The same formulae can be derived from the PKE.

For particles in the spherically symmetric gravitational field,
formula (\ref{eq19}) takes the form
\begin{equation}
\bm o=\bm\Omega-\bm\omega=\frac{m}{\bm p^2}\bigl(\bm g\times\bm
p\bigr). \label{rel}\end{equation}

If the angle of particle momentum deflection $\theta$
is small, the evolution of the helicity
is described by the equation \cite{PRD}
\begin{equation}
\zeta=1-\frac{\theta^2}{2(2\gamma-\gamma^{-1})^2},\label{eq28}\end{equation}
where $\gamma=\epsilon/m$ is the Lorentz factor. The original helicity
is supposed to be equal to $+1$.

The relative angular velocity defining the helicity evolution
in the uniformly accelerated frame is given by
\begin{equation}
\bm
o=\bm\Omega-\bm\omega=-\frac{m}{\bm p^2}\bigl(\bm a\times\bm
p\bigr). \label{eq34}\end{equation}

When $\bm a=-\bm g$, values of $\bm o$ in Eqs. (\ref{eq34}) and (\ref{rel}) are the same. It is the manifestation
of the equivalence principle which was discussed with respect to helicity evolution in  \cite{T1,T2}.

At the same time, the manifestation of the equivalence principle
for the spin rotation is not so trivial. In particular, the spin
of nonrelativistic particles in the spherically symmetric
gravitational field rotates three times more rapidly in comparison
with the corresponding accelerated frame \cite{PRD}.

The helicity evolution caused by the rotation of an astrophysical
object was considered in Ref. \cite{T1}. The effect of the
rotation of a field source is characterized by the gravitomagnetic
field. This field makes the velocity rotate twice faster than the
spin and changes the helicity. Therefore, the helicity can locally
evolve due to the rotation of the field source. Nevertheless, the
integral effect for the particle passing throughout the
gravitational field region is zero. Thus, the helicity of the
scattered massive particle is not influenced by the rotation of an
astrophysical object \cite{T1}. Some other authors came to the
alternative conclusion that the above mentioned rotation changes
the helicity of the scattered massive particle \cite{P1,P2P3}. To
obtain a definite solution of this problem, the PKE and the Dirac
equation can be used.

\section{Discussion and summary}

The general equations describing the dynamics of classical spin in
gravitational fields and noninertial frames was obtained by
Mathisson and Papapetrou \cite{Mathisson,Papapetrou} and by
Pomeransky and Khriplovich \cite{PK}. The MPE and PKE are
different in principle. Nevertheless, the spin dynamics
predicted by the MPE and PKE is the same in the first-order approximation
in the spin. This important conclusion shows that the
Mathisson-Papapetrou and Pomeransky-Khriplovich approaches lead to the same
observable spin effects. Results obtained with two approaches differ only
in terms of the second and higher orders in spin. These terms are
proportional to derivatives of the second and higher orders of the metric
tensor. Both of approaches predict the violation of the weak equivalence
principle due to deflections of spinning particles from geodesic lines.
Such deflections are caused by the gravitational Stern-Gerlach forces
which are rather weak. Nevertheless, these forces are important because
they condition
the violation of the weak equivalence principle \cite{Plyatsko}.
The Mathisson-Papapetrou and Pomeransky-Khriplovich approaches give
different expressions for the gravitational Stern-Gerlach forces.
The expression resulting from the
PKE agrees with that obtained from the Dirac equation.

The PKE are rather convenient for description of spin motion in
the framework of classical gravity. The general equation of spin
motion \cite{PK} is valid in arbitrary spacetimes. However, the
angular velocity of spin precession defined by Eq. (\ref{omega})
depends on the choice of the tetrad. The origin of such a
dependence is the fact that reference frames defined by different
tetrads can move relatively to each other. In this case, the
corresponding angular velocities of spin precession differ due to
the Thomas precession. We derive the exact equation describing the
spin dynamics in the rotating frame.

An important property of spin interactions with curved spacetimes is the
absence of the AGM and GDM \cite{KO}. The relations obtained by Kobzarev
and Okun lead to equal frequencies of precession of classical and
quantum spins in curved spacetimes and the preservation of helicity
of Dirac particles in gravitomagnetic fields \cite{T1}. As a result,
the behavior of classical and quantum spins in curved spacetimes is the same
and any quantum effects cannot appear. However, this point of view was not
generally accepted until very recently.

The fact that dynamics of classical and quantum spins in curved spasetimes
is identical was also proved in Refs. \cite{PRD,PRD2}. The full agreement between
classical equations of momentum and spin motion and corresponding quantum
equations obtained from solution of the Dirac equation was established.
The classical and quantum equations was compared not only for gravitational fields
but also for noninertial frames. The absence of any fundamentally
new spin effects is a manifestation of the correspondence principle.

Another manifestation of the equivalence principle is the helicity
evolution.
While the motion of momentum and spin differs in static
gravitational fields and uniformly accelerated frames, the
helicity evolution is the same \cite{PRD}.

\section*{Acknowledgements}

The author would like to thank the Organizing committee of the
Mat\-his\-son conference and the Stefan Banach Intenational
Mathematical Center for the hospitality and support. The author is
also grateful to O.V. Teryaev for important remarks and helpful
discussions and acknowledges the support by the Belarusian
Republican Foundation for Fundamental Research.


\end{document}